\documentclass{JAC2001}
\usepackage{graphicx}
\newcommand{\ud}{\mathrm{d}}
\begin{document}
\title{THE SHORT-TERM DYNAMICAL APERTURE
VIA VARIATIONAL-WAVELET APPROACH WITH  CONSTRAINTS}
\author{Antonina N. Fedorova,  Michael G. Zeitlin \\ 
IPME, RAS, V.O. Bolshoj pr., 61, 199178, St.~Petersburg, Russia 
\thanks{e-mail: zeitlin@math.ipme.ru}\thanks{ http://www.ipme.ru/zeitlin.html;
http://www.ipme.nw.ru/zeitlin.html}
}

\maketitle

\begin{abstract}
We present the applications of wavelet analysis methods       
in constrained variational framework to calculation of dynamical aperture.
We construct represention  via exact nonlinear high-localized periodic eigenmodes      
expansions, which  allows to control contribution to motion from each scale     
of underlying multiscale structure and consider qualitative approach to the problem.          
\end{abstract}

\section{INTRODUCTION}

The estimation of dynamic aperture of accelerator is real and long standing 
problem. From the formal point of view the aperture is a some border 
between  two types of dynamics: relative regular and predictable motion along 
of acceptable orbits or fluxes of orbits corresponding to KAM tori and stochastic motion 
with particle losses   
blown away by Arnold diffusion and/or chaotic motions.
According to standard point of view this transition is being done 
by some analogues with map technique [1]. 
Consideration for aperture of n-pole Hamiltonians with kicks 

\begin{eqnarray}
&&H=\frac{p_x^2}{2}+\frac{K_x(s)}{2}x^2+\frac{p_y^2}{2}+\frac{K_y(s)}{2}y^2+\\
&&\frac{1}{3!B\rho}\frac{\partial^2B_z}{\partial x^2}(x^3-3xy^2)L\sum_{k=-\infty}^{\infty}
\delta(s-kL)+\dots \nonumber
\end{eqnarray}

is done by linearisation and
discretization of canonical transformation and the result resembles (pure formally) standard mapping.
This leads, by using Chirikov criterion of resonance overlapping,
to evaluation of aperture via amplitude of the following global harmonic 
representation:

\begin{eqnarray}
x^{(n)}(s)&=&\sqrt{2J_{(n)}\beta_x(s)}\cdot\\
&&\cos\Big(\psi_1-\frac{2\pi\nu}{L}s+
\int_0^s \frac{\ud s'}{\beta_x(s')}\Big)\nonumber
\end{eqnarray}

The goal of this paper is is two-fold. In part 2 we consider some qualitative criterion
which is based on more realistic understanding of difference between motion
in KAM regions and stochastic regions: motion in KAM regions may be described only by regular
functions (without rich internal structures) but motion in stochastic regions/layers
may be described by functions with internal self-similar structures, i.e. fractal type functions.
Wavelet analysis approach [2], [3] provides us with more or less analytical description based on
calculations of wavelet coefficients/wavelet transform asymptotics.
In part 3 we consider the same problem on a more quantitative level as constrained variational problem
and give explicit representation for all dynamical variables 
as expansions in nonlinear periodic high-localized eigenmodes.

\section{QUALITATIVE ANALYSIS}

Fractal or chaotic image is a function (distribution), which has structure at all
underlying scales. 
So, such objects have additional nontrivial details on any level of resolution.
But such objects cannot be represented by smooth functions, because they 
resemble constants at small scales [2], [3].
So, we need to find self-similarity behaviour during movement to 
small scales for functions describing non-regular motion.
So, if we look on a ``fractal'' function $f$ (e.g. Weierstrass function) near an
arbitrary point at different scales, we find the same function up to a scaling factor.
Consider the fluctuations of such function $f$ near some point $x_0$
\begin{equation}
f_{loc}(x)=f(x_0+x)-f(x_0)
\end{equation} 
then we have 
\begin{equation}
f_{x_0}(\lambda x)\sim\lambda^{\alpha(x_0)}f_{x_0}(x)
\end{equation} 
where $\alpha(x_0)$ is the local scaling exponent or H\"older exponent of the 
function $f$ at $x_0$.

According to [3] general functional spaces and scales of spaces 
can be characterized through wavelet coefficients or wavelet transforms. 
Let us consider continuous wavelet transform
$$
W_g f(b,a)=\int_{R^n}\ud x\frac{1}{a^n}\bar g\left(\frac{x-b}{a}\right) f(x),
$$
$ b\in R^n, \quad a>0$,
w.r.t. analyzing wavelet $g$, which is strictly admissible, i.e.
$$
C_{g,g}=\int_0^\infty({\ud a}/{a})\vert\bar{\hat g(ak)}\vert^2<\infty
$$
Wavelet transform has the following covariance property under action of underlying affine group:
\begin{eqnarray}
W_g(\lambda a, x_0+\lambda b)\sim\lambda^{\alpha(x_0)}W_g(a,x_0+b)
\end{eqnarray}
So, if H\"older exponent of  (distribution) $f(x)$ around the point $x=x_0$ is 
$h(x_0)\in (n,n+1)$, then we have the following behaviour of $f(x)$ around $x=x_0$:
$
f(x)=c_0+c_1(x-x_0)+\dots+$
$c_n(x-x_0)^n+c|x-x_0|^{h(x_0)}$.
Let analyzing wavelet have $n_1$ ($>n$) vanishing moments, then
\begin{eqnarray}
W_g(f)(x_0,a)=Ca^{h(x_0)}W_g(f)(x_0,a)
\end{eqnarray} 
and $W_g(f)(x_0,a)\sim a^{h(x_0)}$  when $a\to 0$.
But if $f\in C^{\infty}$ at least in point $x_0$, then
$W_g(f)(x_0,a)\sim a^{n_1}$  when $a\to 0$.
This shows that localization of the wavelet coefficients at small
scale is linked to local regularity.
As a rule, the faster
the wavelet coefficients decay, the more the analyzed function is
regular. 
So,transition from regular motion to chaotic one may be characterised as the changing of
H\"older exponent of function, which describes motion.
This gives criterion of appearance of fractal behaviour and may determine,at least in principle,
dynamic aperture.

\section{CONSTRAINED PROBLEM FOR QUASI-PERIODIC ORBITS}

We consider extension of our approach [4]-[15] to the case
of constrained quasi-periodic trajectories. The equations of motion corresponding
to Hamiltonian (1) may be formulated as a particular case of
the general system of
ordinary differential equations
$
{dx_i}/{ds}=f_i(x_j,s)$, $  (i,j=1,...,2n)$, 
where $f_i$ are not more
than rational functions of dynamical variables $x_j$
and  have arbitrary dependence of time but with periodic boundary conditions.
Let us consider this system as an operator equation for operator $S$, which satisfies the equation  
\begin{equation}
S(H,x,\partial/\partial s,\partial/\partial x,s)=0
\end{equation}
which is polynomial/rational in $x=(x_1$, $\dots$ , $x_n$, $ p_1$, $\dots$ , $p_n)$ and have arbitrary 
dependence on $s$ 
and operator $C(H$, $x$, $\partial/\partial t$, $\partial/\partial x$, $s$), which  is an operator describing
some constraints as differential as integral on the set of dynamical variables.
E.g., we may fix a part of non-destroying integrals of motion (e.g., energy) or 
areas in phase space (fluxes of orbits). So, we may consider our problem as constructing   
orbits described by Hamiltonian (1).
In this way we may fix a given acceptable aperture or vice versa by feedback 
via parametrisation of orbits by coefficients of initial dynamical problem we may control different levels
of aperture as a function of the parameters of the system (1) under consideration.
As a result our variational problem is formulated for pair of operators (C, S) 
on extended set of dynamical variables which includes Lagrangian multipliers $\lambda$.

Then we use (weak) variation formulation  
\begin{equation}
\int<(S+\lambda C)x, y>\ud t=0
\end{equation}

We start with hierarchical sequence of approximations spaces:
\begin{eqnarray}
\dots V_{-2}\subset V_{-1}\subset V_{0}\subset V_{1}\subset V_{2}\dots,
\end{eqnarray}
and the corresponding expansions:
\begin{eqnarray}
x^N(s)=\sum^N_{r=1}a_r\psi_r(s), \quad
y^N(s)=\sum^N_{k=1}b_k\psi_k(s)
\end{eqnarray}
As a result
we have from (7) the following reduced system of algebraical equations (RSAE)
on the set of unknown coefficients $a_i$ of
expansions (10):
\begin{eqnarray}\label{eq:pol2}
L(S_{ij},C_{kl},a,\Lambda)=0
\end{eqnarray}
where operator L is algebraization of initial problem (7) and
we need to find
in general situation objects $\Lambda$.
\begin{eqnarray}
\Lambda^{d_1 d_2 ...d_n}_{\ell_1 \ell_2 ...\ell_n}=
 \int\limits_{-\infty}^{\infty}\prod\psi^{d_i}_{\ell_i}(x)\ud x,
\end{eqnarray}
We consider the procedure of their
calculations in case of quasi/periodic boundary conditions
in the bases of periodic wavelet functions with periods $T_i$ on
the interval [0,T] and the corresponding expansion (10) inside our
variational approach. Periodization procedure
gives 
\begin{eqnarray}
\hat\varphi_{j,k}(x)&\equiv&\sum_{\ell\in Z}\varphi_{j,k}(x-\ell)\\
\hat\psi_{j,k}(x)&\equiv&\sum_{\ell\in Z}\psi_{j,k}(x-\ell)\nonumber
\end{eqnarray}
So, $\hat\varphi, \hat\psi$ are periodic functions on the interval
[0,T]. Because $\varphi_{j,k}=\varphi_{j,k'}$ if $k=k'\mathrm{mod}(2^j)$, we
may consider only $0\leq k\leq 2^j$ and as  consequence our
multiresolution has the form
$\displaystyle\bigcup_{j\geq 0} \hat V_j=L^2[0,T]$ with
$\hat V_j= \mathrm{span} \{\hat\varphi_{j,k}\}^{2j-1}_{k=0}$ [16].
Integration by parts and periodicity gives  useful relations between
objects (12) in particular quadratic case $(d=d_1+d_2)$:
\begin{eqnarray}
\Lambda^{d_1,d_2}_{k_1,k_2}&=&(-1)^{d_1}\Lambda^{0,d_2+d_1}_{k_1,k_2},\\
\Lambda^{0,d}_{k_1,k_2}&=&\Lambda^{0,d}_{0,k_2-k_1}\equiv
\Lambda^d_{k_2-k_1}\nonumber
\end{eqnarray}
So, any 2-tuple can be represented by $\Lambda^d_k$.
Then our second (after (11)) additional algebraic (linear) problem is reduced according to [16] to the eigenvalue
problem for
$\{\Lambda^d_k\}_{0\leq k\le 2^j}$ by creating a system of $2^j$
homogeneous relations in $\Lambda^d_k$ and inhomogeneous equations.
So, if we have dilation equation in the form
$\varphi(x)=\sqrt{2}\sum_{k\in Z}h_k\varphi(2x-k)$,
then we have the following homogeneous relations
\begin{equation}
\Lambda^d_k=2^d\sum_{m=0}^{N-1}\sum_{\ell=0}^{N-1}h_m h_\ell
\Lambda^d_{\ell+2k-m},
\end{equation}
or in such form
$A\lambda^d=2^d\lambda^d$, where $\lambda^d=\{\Lambda^d_k\}_
{0\leq k\le 2^j}$.
Inhomogeneous equations are:
\begin{equation}
\sum_{\ell}M_\ell^d\Lambda^d_\ell=d!2^{-j/2},
\end{equation}
 where objects
$M_\ell^d(|\ell|\leq N-2)$ can be computed by recursive procedure
\begin{eqnarray}
&&M_\ell^d=2^{-j(2d+1)/2}\tilde{M_\ell^d},\\ 
&&\tilde{M_\ell^k}=
<x^k,\varphi_{0,\ell}>=
\sum_{j=0}^k {k\choose j} n^{k-j}M_0^j,\quad
\tilde{M_0^\ell}=1.\nonumber
\end{eqnarray}
So, this problem is the standard
linear algebraical problem. 

Then, we may solve RSAE (\ref{eq:pol2}) and determine
unknown coefficients from formal expansion (10) and therefore to 
obtain the solution of our initial problem.
It should be noted that if we consider only truncated expansion with N terms
then we have from (\ref{eq:pol2}) the system of $N\times 2n$ 
algebraical equations 
and the degree of this algebraical system coincides
with the degree of initial differential system.
As a result we obtained
the following explicit representation for periodic trajectories 
in the base of periodized (period $T_i$) wavelets (10):
\begin{eqnarray}
x_i(s)=x_i(0)+\sum_k a_i^k\psi_k^i(s),\quad x_i(0)=x_i(T_i),
\end{eqnarray}
Because affine
group of translation and dilations is inside the approach, this
method resembles the action of a microscope. We have contribution to
final result from each scale of resolution from the whole
infinite scale of spaces. More exactly, the closed subspace
$V_j (j\in {\bf Z})$ corresponds to  level j of resolution, or to scale j.
The solution has the following form
\begin{equation}\label{eq:z}
x(s)=x_N^{slow}(s)+\sum_{j\geq N}x_j(\omega_js), \quad \omega_j\sim 2^j
\end{equation}

\begin{figure}[htb]
\centering
\includegraphics*[width=60mm]{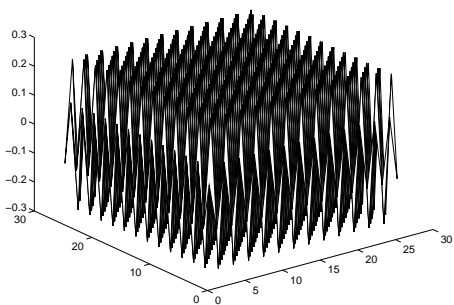}
\caption{Periodic structure on level 6.}
\end{figure}

which corresponds to the full multiresolution expansion in all time 
scales.
Formula (\ref{eq:z}) gives us expansion into a slow part $x_N^{slow}$
and fast oscillating parts for arbitrary N. So, we may move
from coarse scales of resolution to the 
finest one for obtaining more detailed information about our dynamical process.
The first term in the RHS of equation (19) corresponds on the global level
of function space decomposition to  resolution space and the second one
to detail space. In this way we give contribution to our full solution
from each scale of resolution or each time scale.
On Fig.~1  we present (quasi) periodic regime on section $x-p_x$ corresponding to model (1).

\section{ACKNOWLEDGMENTS}

We would like to thank The U.S. Civilian Research \& Development Foundation (CRDF) for
support (Grants TGP-454, 455), which gave us the possibility to present our nine papers during
PAC2001 Conference in Chicago and Ms. Camille de Walder from CRDF for her help and encouragement.

\end{document}